\documentclass[iop]{emulateapj}
\pdfoutput=1
\usepackage{natbib}
\usepackage{threeparttable}
\usepackage{amsmath}
\usepackage{amssymb}

\shortauthors{Vagnozzi et al.}

\begin{document}

\title{Solar models in light of new high metallicity measurements from solar wind data}

\author{
\mbox{Sunny Vagnozzi{$^{1}$},}
\mbox{Katherine Freese{$^{1,2}$},}
and
\mbox{Thomas H. Zurbuchen{$^{3,4}$}}
}

\affiliation{
\mbox{$^1$ The Oskar Klein Centre for Cosmoparticle Physics, Stockholm University, SE-106 91 Stockholm, Sweden}
\mbox{$^2$ Michigan Center for Theoretical Physics, Department of Physics, University of Michigan, Ann Arbor, MI 48109, USA} 
\mbox{$^3$ NASA Headquarters, 300 E St SW, Washington, DC 20546, USA}
\mbox{$^4$ Department of Climate and Space Sciences and Engineering, University of Michigan, Ann Arbor, MI 48109, USA}
}

\email{sunny.vagnozzi@fysik.su.se}
\email{ktfreese@umich.edu}
\email{thomas.h.zurbuchen@nasa.gov}

\begin{abstract}
We study the impact of new metallicity measurements, from solar wind data, on the solar model. The ``solar modelling problem" refers to the persisting discrepancy between helioseismological observations and predictions of solar models computed implementing state-of-the-art photospheric abundances. We critically reassess the problem, in particular considering the new set of abundances of von Steiger \& Zurbuchen 2016, determined through the \textit{in situ} collection of solar wind samples from polar coronal holes. This new set of abundances indicates a solar metallicity $Z_{\odot} \geq 0.0196 \pm 0.0014$, significantly higher than the currently established value. The new values hint at an abundance of volatile elements (i.e. C, N, O, Ne) close to previous results of Grevesse \& Sauval 1998, whereas the abundance of refractory elements (i.e. Mg, Si, S, Fe) is considerably increased. Using the Linear Solar Model formalism, we determine the variation of helioseismological observables in response to the changes in elemental abundances, in order to explore the consistency of these new measurements with constraints from helioseismology. We find that, for observables that are particularly sensitive to the abundance of volatile elements, in particular the radius of the convective zone boundary (CZB) and the sound speed around the radius of CZB, improved agreement over previous models is obtained. Conversely, the high abundance of refractories correlates with a higher core temperature, resulting in an overproduction of neutrinos and a huge increase in the surface Helium abundance. We conclude that the ``solar modelling problem" remains unsolved.
\end{abstract}


\section{\textbf{Introduction}}\label{sec:introduction}

A major issue in solar physics, known as the ``solar modelling problem", has emerged over the past decade, following a significant systematic downward revision of solar metallicity (Asplund et al. 2005; Asplund et al. 2009: Caffau et al. 2011; Grevesse et al. 2015; Scott et al. 2015a; Scott et al. 2015b). Standard Solar Models (SSM) constructed with these heavy element mixtures are in apparent conflict with helioseismic probes of the solar interior, which include the sound speed profile, the radius of the convective zone boundary (CZB) and the surface Helium abundance (for reviews, see e.g. Serenelli et al. 2009). For instance, the sound speed is inferred to be $\sim 1\%$ lower than predicted at the radius of CZB. Similarly, the surface Helium abundance and the radius of CZB are $\sim 7\%$ lower and $\sim 1.5\%$ higher than those deduced from helioseismology (see Villante 2010 for the quoted numbers). Given the precision at which we are capable of measuring helioseismological observables, the above represent discrepancies of the order of several $\sigma$s (see e.g. Villante 2015).

There has been no shortage of proposed solutions, which include an anomalously large Ne abundance in the photosphere (Bahcall et al. 2005), physical processes not accounted for in the SSM (Montalban et al. 2004; Guzik et al. 2005; Drake \& Testa 2005; Charbonnel \& Talon 2005; Castro et al. 2007; Guzick \& Mussack 2010; Turck-Chieze et al. 2010; Turck-Chieze et al. 2011; Serenelli et al. 2011; Yang 2016), axion-like particles (Vincent et al. 2013), missing opacity (Christensen-Dalsgaard et al. 2009; Serenelli et al. 2009; Villante \& Ricci 2010; Villante 2010; Villante et al. 2014; Villante \& Serenelli 2015), and finally, exotic energy transport by captured dark matter (Frandsen \& Sarkar 2010; Cumberbatch et al. 2010; Taoso et al. 2010; Lopes et al. 2014; Vincent et al. 2015a; Vincent et al. 2015b; Dev \& Teresi 2016; Vincent et al. 2016; Geytenbeek et al. 2016). However, none of these ideas seem to adequately solve the ``solar modelling problem" (see e.g. Shearer et al. 2014).

In this paper we will instead investigate the possibility that the metallicity of the Sun may not be sufficiently well known. The aforementioned \textit{low-Z} metallicity measurements rely on the methodology of photospheric spectroscopy. Our approach is motivated by a completely different technique to estimate photospheric abundances, based on \textit{in situ} measurements of heavy ions in the least fractionated solar wind accessible for direct \textit{in situ} study of the photosphere. In particular von Steiger \& Zurbuchen 2016 (vSZ16 henceforth), adopting the solar wind methodology, determine a value for the solar metallicity which is significantly higher than suggested by spectroscopic estimates.

Our paper does not seek to take sides between the two different methodologies to calculate the solar metallicity, but instead to estimate the consequences of the vSZ16 methodology on solar models. In order to do this, we make use of the \textit{Linear Solar Model} (LSM) formalism introduced in Villante \& Ricci 2010 to explore the consistency of these new abundance measurements with constraints from helioseismology.

The remainder of the paper is organized as follows. Section~\ref{sec:solarwind} discusses more in depth \textit{in situ} solar wind measurements of solar metallicity. Section~\ref{sec:method} will provide details on the methodology adopted to study the impact on helioseismology observables. In Section~\ref{sec:results} we present our results, as well a caveat to the applicability of our methodology. In Section~\ref{sec:conclusion} we discuss the implications of these results for helioseismology, and provide concluding remarks.

\section{\textbf{\textit{In situ} solar wind measurements of metallicity}}\label{sec:solarwind}

There are a number of different ways by means of which solar metallicity, $Z_{\odot}$, can be measured, and none of them is simple or straightforward. The aforementioned \textit{low-Z} abundance catalogues, in particular that of Asplund et al. 2009 (AGSS09 henceforward), have been compiled making use of the methods of photospheric spectroscopy. Despite its broad use within the solar physics community, spectroscopy is not immune to drawbacks and systematics. The interpretation of its observations requires sophisticated forward modelling techniques which account for radiative transport, three-dimensional structure and hydrodynamic models of the observation volumes, and departures from local thermodynamic equilibrium. In addition, the methodology also relies on detailed knowledge of the relevant atomic and molecular transition probabilities.

An alternative method for determining the solar metallicity relies instead on \textit{in situ} collection of solar samples, which eliminates the need for forward modelling, but adds possible fractionation effects when \textit{in situ} measurements are to be used to constrain the solar metallicity. For solar wind plasma compositions, various processes in the low coronal can affect the abundance of ions, based on their ionization, gravitational settling and transport histories. Examples of fractionation processes at work are collisional coupling (especially for He), First Ionization Potential (FIP) fractionation (Hovestadt et al. 1973; Bochsler 2000) presumably operating in the low solar atmosphere, and gravitational settling (Geiss et al. 1970; Weberg et al. 2012).

Among all solar wind samples in the heliosphere, solar wind from polar coronal holes (PCHs) is the least fractionated of all samples of steady state of transient solar wind flows (Zurbuchen 2007; Zurbuchen et al. 2012; Zurbuchen et al. 2016). Even in PCH-associated wind, there is still some fractionation, especially of insufficient collisional coupling that affects all of solar wind, but is most evident in He/H (Geiss et al. 1970). Furthermore, it was shown that the composition of these PCHs is constant during the entire \textit{Ulysses} mission (Smith et al. 1992), which explored these polar regions during the period from 1990 to 2009. The only observed residual changes relate to small variations of the ionization state, reflecting the temperature and acceleration history of the solar wind emerging from PCHs. The elemental composition remains constant within the error bars of this methodology (McComas et al. 2008; von Steiger \& Zurbuchen 2011).

As discussed in von Steiger \& Zurbuchen 2016, any other residual fractionation in these regions cannot be excluded. However, based on the physical processes and a systematic study of various source regions, they concluded that such processes systematically decrease the overall solar metallicity (see von Steiger \& Zurbuchen 2015, Zurbuchen et al. 2016, for details). If this is correct, the \textit{in situ} measured $Z_{\odot}$ represents a lower limit to the true metallicity of the Sun. 

It is worth pointing out why these measurements were only recently published. Previous data inversion methodologies used long-term averages and statistical inversion techniques as discussed by von Steiger et al. 2000. Only recently Shearer et al. 2014 generalized the inversion techniques to low count rates and leading to statistically robust estimates of trace elements used for the $Z_{\odot}$ measurement. Based on this analysis, vSZ16 reports a lower limit on the solar metallicity, $Z_{\odot} \geq 0.0196 \pm 0.0014$, which is significantly higher than the widely used AGSS09 value of $Z_{\odot} = 0.0133$ (which we take as our baseline model from here on). The sample analyzed has been shown to be most representative of that of the photosphere, in contrast to low-latitude and/or transient solar wind which is more prone to fractionation (Feldman et al. 1998; Reisenfeld et al. 2013; von Steiger \& Zurbuchen 2016, Zurbuchen et al. 2016).

The value derived by vSZ16 is significantly closer to previous \textit{high-Z} compositions of Anders \& Grevesse 1989, and Grevesse \& Sauval 1998 (GS98), which preceded the aforementioned downward revision of metallicity and which also yielded reasonable agreement with helioseismology. However, although the total metallicity is similar, the details concerning individual elemental abundances (in particular, the abundance of refractory elements) are quite different, an aspect which will have very important consequences for our subsequent considerations.

\section{\textbf{Method}}\label{sec:method}

Our goal is to provide a first inspection of solar models in light of the \textit{high-Z} composition presented by vSZ16, in particular whether the new composition can restore consistency with helioseismology. Here, we shall limit ourselves to conducting a first-order analysis of the problem making use of the \textit{Linear Solar Model} (LSM) methodology, developed in Villante \& Ricci 2010, and Villante 2010. We expect our simple semi-analytical approach to provide useful insight into the behavior of helioseismological observables in response to the change in composition being considered, but leave sophisticated numerical treatments to future work. Furthermore, we note that our results agree with those obtained using a full non-linear treatment in Serenelli et al. 2016, confirming \textit{a posteriori} the goodness of our linear analysis.

\subsection{The role of opacity in solar models}\label{sec:opacity}

The ``solar modelling problem" is deeply rooted to the role played by radiative opacity, $\kappa(r)$, in the SSM. Opacity is a key quantity which describes the tight coupling between radiation and matter in the hot dense interior of the Sun. The main contributor to the opacity profile of the Sun is constituted by metals, which contribute to the opacity through physical processes such as absorption by photoexcitation and photoionization.

Variations of the metal content of the Sun can be effectively described as a fractional variation in its opacity profile, $\delta \kappa (r)$ [to be defined more precisely in Eq.(\ref{defdk})]. Let us take a baseline model of the Sun with abundances $\{Z_i\}$. Consider, then, a variation in abundances $\{Z_i\} \rightarrow \{\overline{Z}_i\}$. The fractional variation in opacity $\delta \kappa(r)$ with respect to the baseline model is defined as follows:
\begin{eqnarray}
\delta \kappa(r) \equiv \frac{\kappa(\overline{Z}_i)}{\kappa(Z_i)}-1 \, .
\label{defdk}
\end{eqnarray}
Therefore, the response of helioseismology observables to abundance variations can be related to the response of the fractional variation in opacity to the same changes.

\begin{figure}[!h]
\includegraphics[width=.3\textwidth , angle=270]{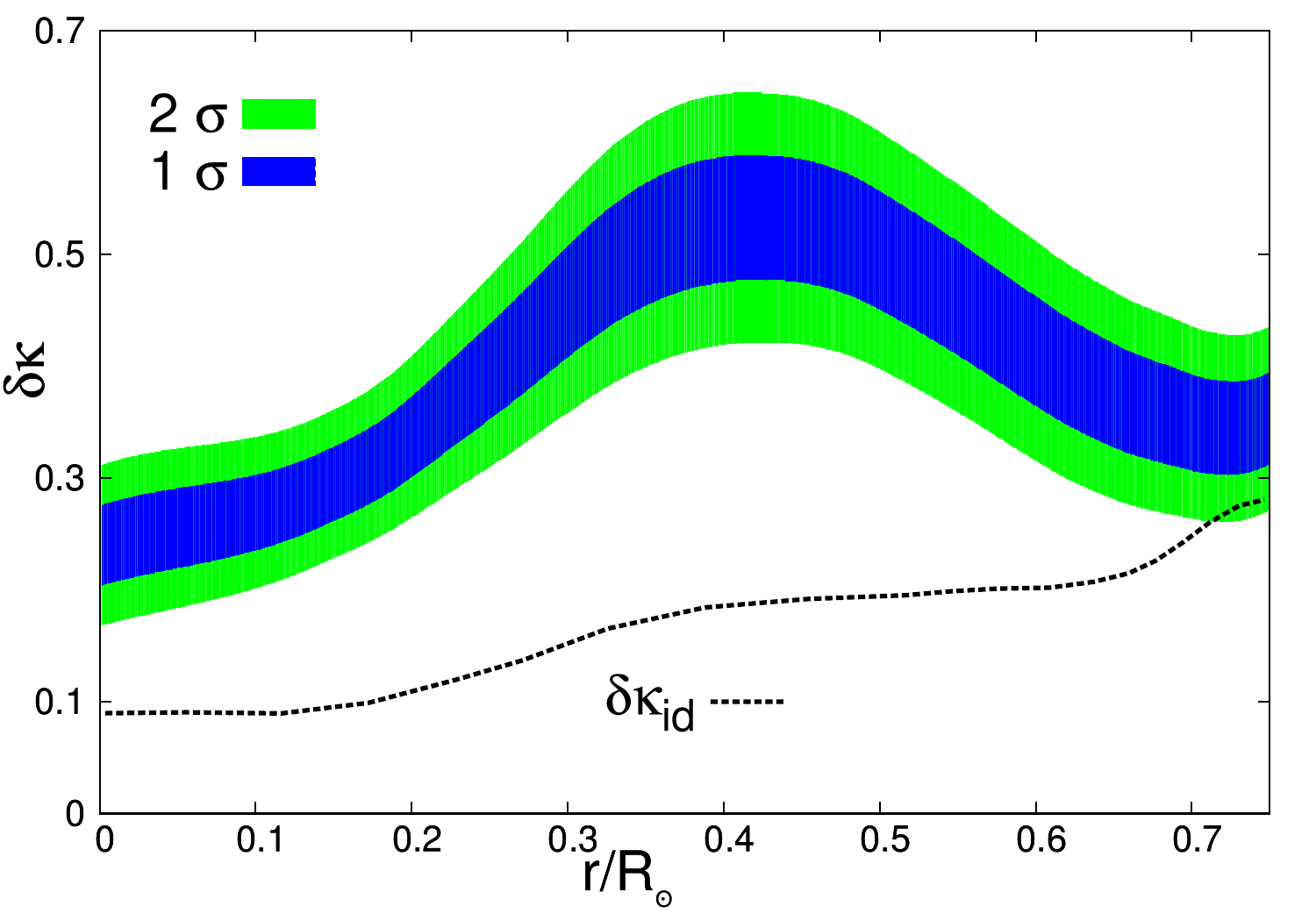}
\caption{Fractional variation in opacity $\delta \kappa (r)$ when comparing the vSZ16 abundances to the baseline AGSS09 abundances. Blue and green bands denote 1$\sigma$ and 2$\sigma$ uncertainty bands, propagated from the uncertainty in the vSZ16 abundances through Eq.~(\ref{deltakappa}). The dashed line denotes the ``ideal" opacity variation $\delta \kappa_{\text{id}}(r)$ with respect to the AGSS09 model which would solve the ``solar modelling problem" while satisfying constraints on the solar neutrino fluxes. The $\delta \kappa_{\text{id}}(r)$ profile we show has been obtained in Villante 2010.}
\label{fig1}
\end{figure}

Recent works have determined that a monotonic approximately linear fractional variation in the opacity with respect to the baseline AGSS09 model, from $\sim 10\%$ near the core to $\sim 30\%$ around the radius of CZB, can restore agreement with helioseismological observables while satisfying constraints from neutrino fluxes (Christensen-Dalsgaard et al. 2009; Serenelli et al. 2009; Villante \& Ricci 2010; Villante 2010; Villante et al. 2014; Villante \& Serenelli 2015). Let us refer to this ``ideal" variation in opacity with respect to the baseline AGSS09 model as $\delta \kappa_{\text{id}}(r)$, which is given by the dashed line in Fig.~\ref{fig1} (we show the $\delta \kappa_{\text{id}}(r)$ profile obtained in Villante 2010). More recent work seems to suggest that the radiative opacity of the Sun is likely to have been underestimated, with a more accurate treatment of effects such as line broadening possibly going in the direction required to address discrepancies that are, at least in part, related to the ``solar modelling problem" (Bailey et al. 2015, Krief et al. 2016).

Following Villante et al. 2014, we express the fractional variation in opacity due to a variation in elemental abundances as:
\begin{eqnarray}
\delta \kappa (r) \simeq  \sum _j \kappa _j(r)\delta Z_j 
\label{deltakappa}
\end{eqnarray}
where $\kappa _j(r)$ is the logarithmic derivative of opacity with respect to metal abundance $Z_j$, that is:
\begin{eqnarray}
\kappa _j(r) \equiv \frac{\partial \ln \kappa (r)}{\partial \ln Z_j}  \, .
\label{kappaj}
\end{eqnarray}
The index $j$ runs over the eight metals contributing to more than 98\% of the metallicity of the Sun: C, O, N, Ne, Mg, Si, S, Fe. By $\delta Z_j$ we denote the fractional variation in the abundance of element $j$ in vSZ16 with respect to its AGSS09 baseline value [we will define $\delta Z_j$ precisely in the next paragraph, see Eq.~(\ref{aidefinition}) and Eq.~(\ref{deltazdefinition})].

Let us provide a formal and operative definition of the fractional variation in elemental abundance of the $j$th element, $\delta Z_j$. To begin with, we define $N_i$ and $N_H$ to be the number of atoms of the $i$th element and Hydrogen which are present in the Sun respectively (from here on, the subscript $_H$ will always refer to Hydrogen). Then, the logarithmic abundance of the $i$th elements relative to Hydrogen, $A_i$, is defined through the following relation:
\begin{eqnarray}
A_i \equiv \log_{10} \frac{N_i}{N_H} + 12 \, .
\label{aidefinition}
\end{eqnarray}
More precisely, $A_i$ correponds to the base 10 logarithm of the number of atoms of the $i$th element for every $10^{12}$ atoms of Hydrogen in the Sun. For simplicity, $A_i$ is usually referred to simply as the logarithmic abundance of the $i$th element, and we will conform to this standard. Notice that, by construction, $A_H = 12$.~\footnote{This is a standard normalization in stellar physics. The motivation behind the choice of the number 12 is that the abundance of some of the rarest elements in the Sun (such as Uranium, Rhenium, Thorium) is of order 1 atom per $10^{12}$ Hydrogen atoms. In this way, the addition of the factor 12 prevents the need for negative numbers, which used to be computationally problematic, when dealing with logarithmic abundances.} Then, given an element $i$ with logarithmic abundances $A_{_{\text{AGSS09}},i}$ and $A_{_{\text{vSZ16}},i}$ according to the AGSS09 and vSZ16 abundances respectively, the fractional variation $\delta Z_i$ [which enters Eq.~(\ref{deltakappa})] can be expressed as:
\begin{eqnarray}
\delta Z_i = 10^{(A_{_{\text{vSZ16}},i}-A_{_{\text{AGSS09}},i})}-1 \, .
\label{deltazdefinition}
\end{eqnarray}
\newline

The abundances of the eight metals according to vSZ16 and AGSS09 are listed in Table~\ref{tab1}, and the variations in their abundances $\delta Z_i$ have been calculated accordingly to Eq.~(\ref{deltazdefinition}). The uncertainties on the vSZ16 abundances have been estimated as 20\% systematics according to Shearer et al. 2014. Notice that the uncertainties on the solar wind measured metallicity values are typically a factor of 2 larger than the corresponding spectroscopic measurements. As can be seen, for all elements other than Ne, the abundances obtained \textit{in situ} are significantly higher, with typical variations of order 0.2 dex or larger. This fact is particularly true for the refractory elements (i.e. Mg, Si, S, Fe), which  crucially will affect all our results (the abundances of refractories in AGSS09 are instead closer to the previous concordance values of GS98). The values of abundances for the volatile elements in vSZ16 (i.e. C, N, O, Ne) are close to the original values of GS98 (which yielded reasonable agreement with helioseismology) especially with regards to C and O.

\begin{figure}[!h]
\includegraphics[width=.3\textwidth , angle=270]{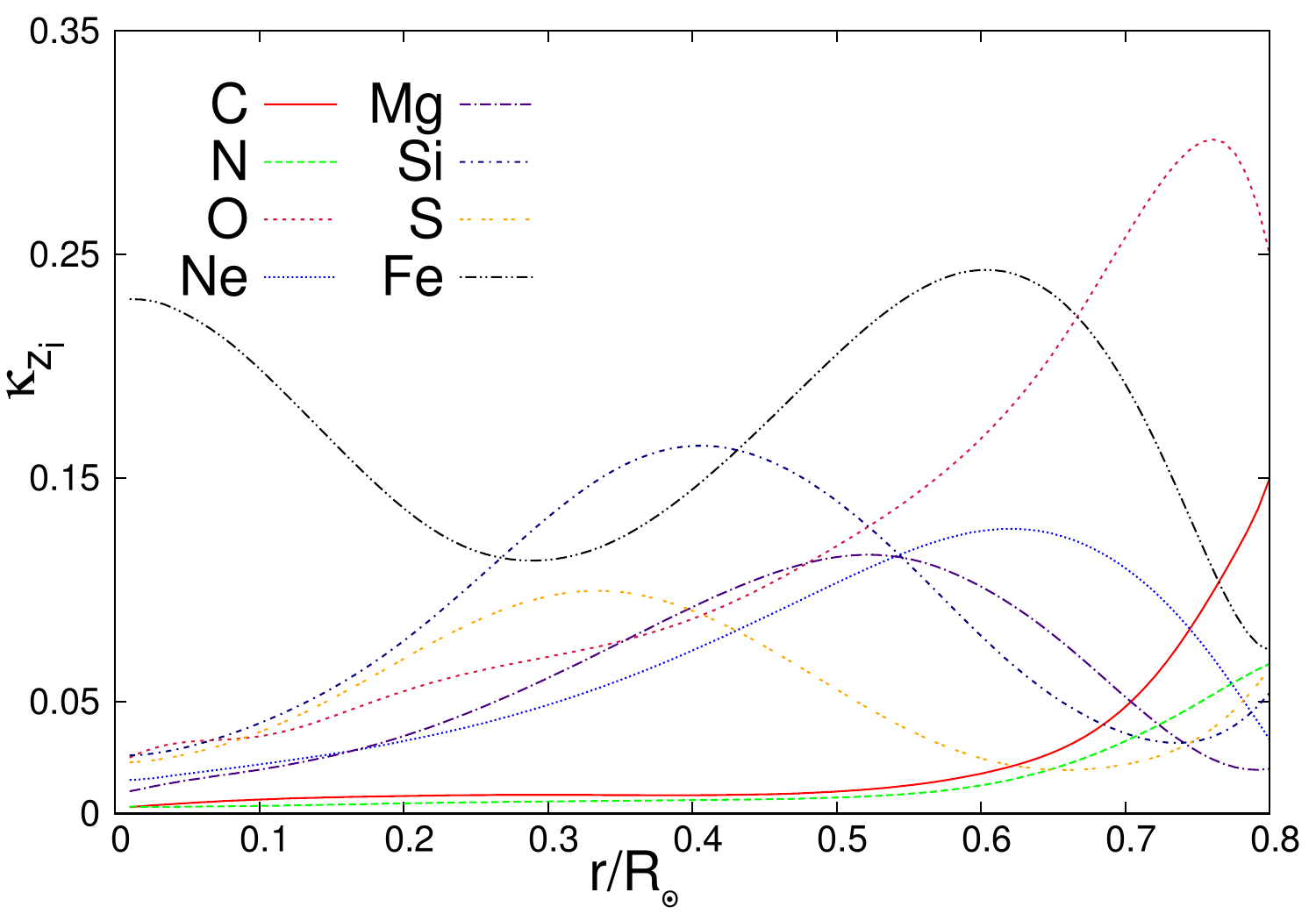}
\caption{Logarithmic derivatives of opacity with respect to individual metal abundances.}
\vskip 0.5 cm
\label{fig2}
\end{figure}

The functional forms of the $\kappa_i$s, i.e. the logarithmic derivatives of radiative opacity with respect to metal abundances, are given in Villante et al. 2014, and plotted in Fig.~\ref{fig2}. We then use Eq.~(\ref{deltakappa}) to estimate the fractional variation in opacity, $\delta \kappa (r)$, associated to the variations in elemental abundances from AGSS09 to vSZ16 listed in Table~\ref{tab1} [this corresponds to the quantity defined in Eq~(\ref{defdk}) when identifying $\{Z_i\}$ and $\{\overline{Z}_i\}$ with the AGSS09 and vSZ16 abundances respectively]. The result is shown in Fig.~\ref{fig1}, including uncertainties propagated by those on the vSZ16 abundances following Eq.~(\ref{deltakappa}). In the same plot, we also compare our profile of opacity variation $\delta \kappa(r)$ with the ``ideal" fractional variation in the opacity with respect to the baseline AGSS09 model, $\delta \kappa_{\text{id}}(r)$. The profile $\delta \kappa_{\text{id}}(r)$ is given by the dashed line, and we notice that it differs substantially from the fractional variation in opacity when going from AGSS09 to vSZ16 abundances we determined, $\delta \kappa(r)$.
\begin{table}[!t]
\caption{Elemental abundances for the AGSS09 and vSZ16 catalogues, and fractional variation between the two.}
\label{tab1}
\begin{tabular}{c c c c}
\hline \hline \\
 \ \ \ Element \ \ \ & \ \ \ $A_{_{\text{AGSS09}}}$ \ \ \ & $A_{_{\text{vSZ16}}}$ \ \ \ & \ \ \ $\delta Z_i$ \ \ \ \\      
\hline
 \ \ \ C \ \ \ & \ \ \ $8.43 \pm 0.05$ \ \ \ & \ \ \ $8.65 \pm 0.08$ \ \ \ & \ \ \ $0.66 \pm 0.15$ \ \ \ \\
 \ \ \ N  \ \ \ & \ \ \ $7.83 \pm 0.05$ \ \ \ & \ \ \ $7.97 \pm 0.08$ \ \ \ & \ \ \ $0.38 \pm 0.08$ \ \ \ \\
 \ \ \ O \ \ \ & \ \ \ $\mathbf{8.69 \pm 0.07}$ \ \ \ & \ \ \ $8.82 \pm 0.11$ \ \ \ & \ \ \ $0.35 \pm 0.10$ \ \ \ \\
 \ \ \ Ne \ \ \ & \ \ \ $7.93 \pm 0.10$ \ \ \ & \ \ \ $7.79 \pm 0.08$ \ \ \ & \ \ \ $-0.28 \pm 0.08$ \ \ \ \\
 \ \ \ Mg \ \ \ & \ \ \ $7.60 \pm 0.04$ \ \ \ & \ \ \ $7.85 \pm 0.08$ \ \ \ & \ \ \ $0.78 \pm 0.16$ \ \ \ \\
 \ \ \ Si \ \ \ & \ \ \ $7.51 \pm 0.03$ \ \ \ & \ \ \ $7.82 \pm 0.08$ \ \ \ & \ \ \ $1.04 \pm 0.21$ \ \ \ \\
 \ \ \ S \ \ \ & \ \ \ $7.12 \pm 0.03$ \ \ \ & \ \ \ $7.56 \pm 0.08$ \ \ \ & \ \ \ $1.75 \pm 0.35$ \ \ \ \\
 \ \ \ Fe \ \ \ & \ \ \ $7.50 \pm 0.04$ \ \ \ & \ \ \ $7.73 \pm 0.08$ \ \ \ & \ \ \ $0.70 \pm 0.15$ \ \ \ \\
\hline
\end{tabular}
\end{table}

Two considerations are in order at this point. The first is that the functional form of $\delta \kappa (r)$ in Fig.~\ref{fig1} (which is principally driven by the large variations in the abundance of two refractory elements, Si and S) differs from the ``ideal" opacity variation with respect to the AGSS09 baseline model $\delta \kappa_{\text{id}}(r)$ we mentioned earlier. Recall $\delta \kappa_{\text{id}}(r)$ consists in a monotonically increasing approximately linear function ranging from $\sim 10\%$ in the core to $\sim 30\%$ at the radius of CZB, and is the variation in opacity required to restore agreement with helioseismology while simultaneously satisfying constraints from solar neutrino fluxes. A look at Fig.~\ref{fig1} reveals how the scale of the $\delta \kappa (r)$ associated to the vSZ16 abundances is larger than the ``ideal" variation $\delta \kappa_{\text{id}}(r)$, represented by the dashed line, by more than 2$\sigma$ over most of the profile of the Sun. Thus, we can already anticipate that the vSZ16 abundances cannot solve the ``solar modelling problem".

The second consideration relates to the observation, already mentioned earlier, that the vSZ16 abundances exhibit a quite contrasting behaviour depending on whether we are considering volatile (i.e. C, N, O, Ne) or refractory (i.e. Mg, Si, S, Fe) elements. These two classes of elements impact different regions of the solar interior: whereas volatiles play a major role around the radius of CZB, refractories strongly impact the conditions in the core. In particular, an increase in the abundance of refractories correlates with a hotter core. The underlying reason is that refractory elements, because of their atomic number (and hence the number number of protons in their nuclei) being higher than that of volatile ones, are able to retain their outer shell electrons bound even in the higher temperatures present in the core. This allows them to make an important contribution to the opacity in the core of the Sun through bound-bound, bound-free, and free-free absorption processes). The increase in opacity makes it harder for photons to escape the core, which thus becomes hotter. The fact that refractories have a large impact on the opacity in the core can be seen by inspecting the kernels $\kappa_{\text{Mg}}$, $\kappa_{\text{Si}}$, $\kappa_{\text{S}}$, $\kappa_{\text{Fe}}$ in Fig.~\ref{fig2}.

As we will discuss more thoroughly in Sec.~\ref{sec:observables}, different helioseismology observables are most sensitive to different regions of the solar interior. Observables which are most sensitive to the opacity and physical conditions around the radius of CZB (such as sound speed around the radius of CZB, as well as the radius of CZB itself, as we will explain subsequently in Sec.~\ref{sec:soundobservables} and Sec.~\ref{sec:czbobservables}), are consequently most sensitive to the abundance of volatiles, and are those for which we can reasonably expect an improvement over AGSS09. Helioseismology observables which instead depend most strongly on the opacity and physical conditions in the solar core (such as surface Helium abundance and neutrino fluxes, as well as the sound speed in the deep interior of the Sun, as we will elucidate in Sec.~\ref{sec:heliumobservables} and Sec.~\ref{sec:neutrinosobservables}) are therefore most sensitive to the abundance of refractories, and are those for which we can expect a worsening over AGSS09. For more thorough discussions on the different impact of volatile and refractory elements on the properties of the Sun, we refer the reader to Serenelli \& Basu 2010, Villante et al. 2014, Villante 2015, Villante \& Serenelli 2015, and Serenelli et al. 2016.

\subsection{Helioseismology observables}\label{sec:observables}

The fractional variation in opacity $\delta \kappa (r)$ of vSZ16 with respect to the baseline AGSS09 model, which we show in Fig.~\ref{fig1}, is used to compute the response of helioseismology observables. We consider four observables: the sound speed $c(r)$, the surface Helium abundance $Y_s$, the radius of CZB $R_b$, and five different solar neutrino fluxes: $\Phi_{\text{pp}}$, $\Phi_{\text{Be}}$, $\Phi_{\text{B}}$, $\Phi_{\text{N}}$, $\Phi_{\text{O}}$.\footnote{We do not include small frequency separation ratios in our analysis because the current formulation of the LSM does not allow us to calculate their response. Moreover, these ratios are strongly correlated with the sound speed, so it would not be correct to adopt both sound speed profile and small frequency separation ratios. Because of this, our results for the sound speed are not directly comparable to the analogous results of Serenelli et al. 2016.}

The idea behind the 	LSM is that, for $\delta \kappa (r) < 1$, the response of the Sun is to good approximation linear in the input variables of the Solar Model (that is, elemental abundances or, equivalently, opacity). Therefore, the fractional variation of a generic given quantity $Q$, $\delta Q \equiv Q/\overline{Q}-1$ (where $\overline{Q}$ is the value of $Q$ in the baseline model), can be related to the fractional opacity variation $\delta \kappa (r)$ through a kernel $K_Q(r)$ as follows:
\begin{eqnarray}
\delta Q = \int dr \ K_Q(r)\delta \kappa (r) \, .
\label{dq}
\end{eqnarray}
\newline

Combining Eq.~(\ref{deltakappa}) and Eq.~(\ref{dq}) it follows that the variation of a generic quantity, $\delta Q$, can be related to the variations in elemental abundances $\delta Z_i$ through power-law exponents ${\cal Q}_i$ as follows (see e.g. Bahcall 1989 for more thorough discussions on power-law exponents):
\begin{eqnarray}
\delta Q(r) = \int dr \ K_Q(r)\sum_i \kappa_i(r)\delta Z_i \equiv \sum_i {\cal Q}_i \delta Z_i \, ,
\label{dq2}
\end{eqnarray}
where the power-law exponents are given by:
\begin{eqnarray}
{\cal Q}_i \equiv \int dr \ K_Q(r)\kappa_i(r) \, .
\label{exponents}
\end{eqnarray}
Recall that the $\kappa_i(r)$ are defined in Eq.~(\ref{kappaj}).

The above Eq.~(\ref{dq2}) and Eq.~(\ref{exponents}) will be useful in propagating uncertainties from the vSZ16 abundances to the final variations in helioseismological observables when going from AGSS09 to vSZ16 abundances. Power-law exponents, moreover, are very useful in understanding the dependence of each helioseismological observable on individual elemental abundances, and in particular whether each observable is most sensitive to the abundance of volatile or refractory elements. We have verified that the power-law exponents recovered as in Eq.~(\ref{exponents}) agree with those tabulated in Villante et al. 2014.

\subsubsection{Sound speed}\label{sec:soundobservables}

\begin{figure}[!h]
\includegraphics[width=.3\textwidth , angle=270]{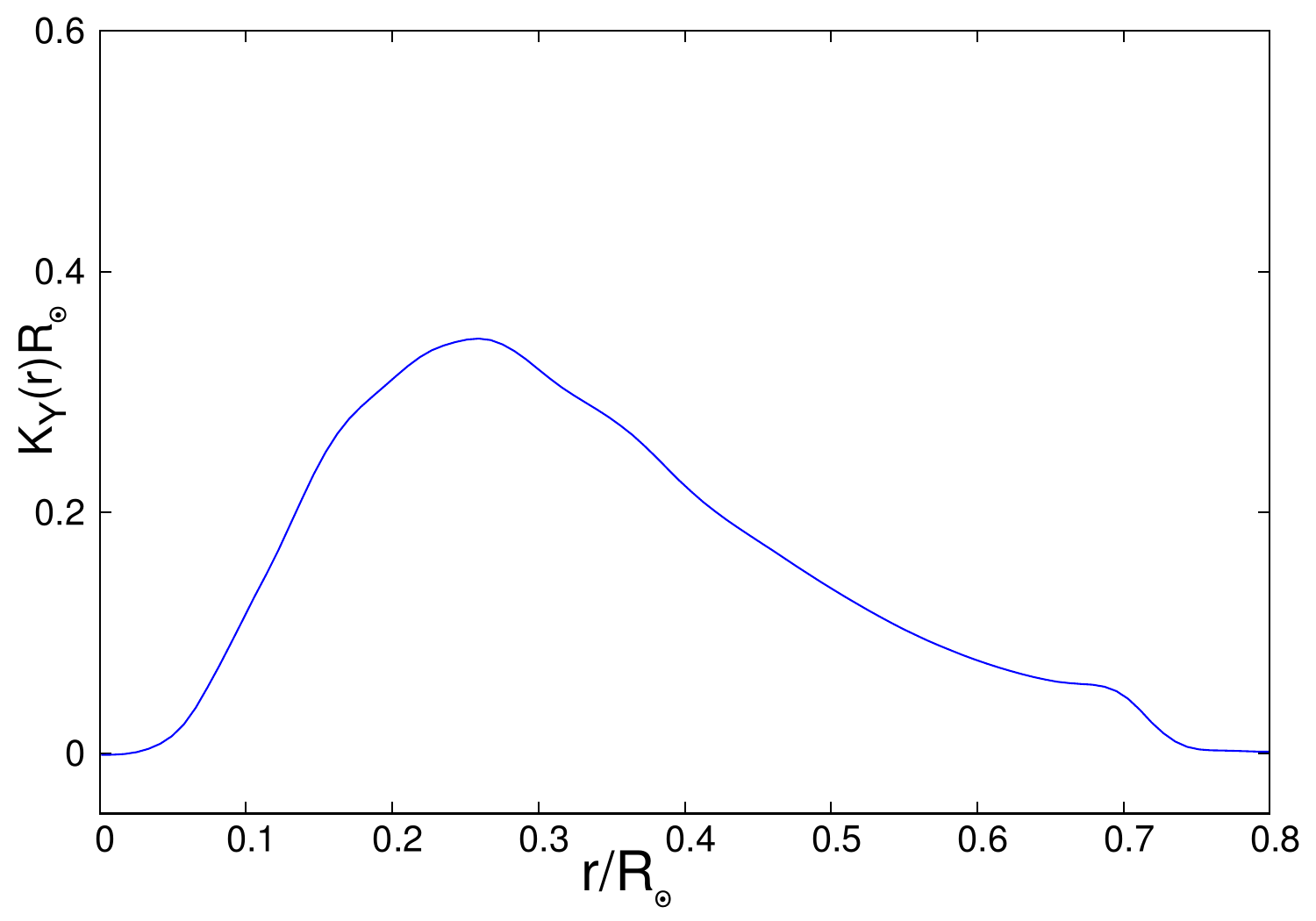}
\caption{Logarithmic derivatives of sound speed with respect to individual metal abundances [see Eq.~(\ref{deltaur})].}
\vskip 0.5 cm
\label{fig3}
\end{figure}

The sound speed kernels, $K_c(r,r')$, have been worked out in Villante 2010. For our purposes, however, it is of more immediate use to consider the logarithmic derivatives of the sound speed with respect to the elemental abundances. These have been calculated in Villante et al. 2014 using the LSM formalism and are shown in Fig.~\ref{fig3}. Here the response of the sound speed $\delta c(r)$ is treated as:
\begin{eqnarray}
\delta c(r) \simeq \sum _j \frac{\partial \ln c(r)}{\partial \ln Z_j}\delta Z_j \equiv \sum _j c_j(r)\delta Z_j \, ,
\label{deltaur}
\end{eqnarray}
where  $c_i(r)$ denotes the logarithmic derivative of the sound speed with respect to the abundance of the $i$th element. 

The sound speed is very sensitive to the opacity profile at the base of the convective zone. This is particularly true for the value of the sound speed at the radius of CZB ($r \approx 0.73 R_{\odot}$), where the predictions of AGSS09 are most discrepant with respect to observations (the sound speed predicted by AGSS09 at that point is too low by $\approx 1\%$). As explained previously, volatile elements play a major part in shaping the opacity profile in that region. In particular, a key role is that played by Oxygen. The abundances of vSZ16 volatiles, in particular that of C and O, are significantly closer to previous concordance values of GS98, than those of AGSS09 are. For this reason, we expect the sound speed profile of vSZ16 to match observations better than that of AGSS09, at least near the convective zone boundary, where AGSS09 was previously most discrepant.

\subsubsection{Surface Helium abundance}\label{sec:heliumobservables}

\begin{figure}[!h]
\includegraphics[width=.3\textwidth , angle=270]{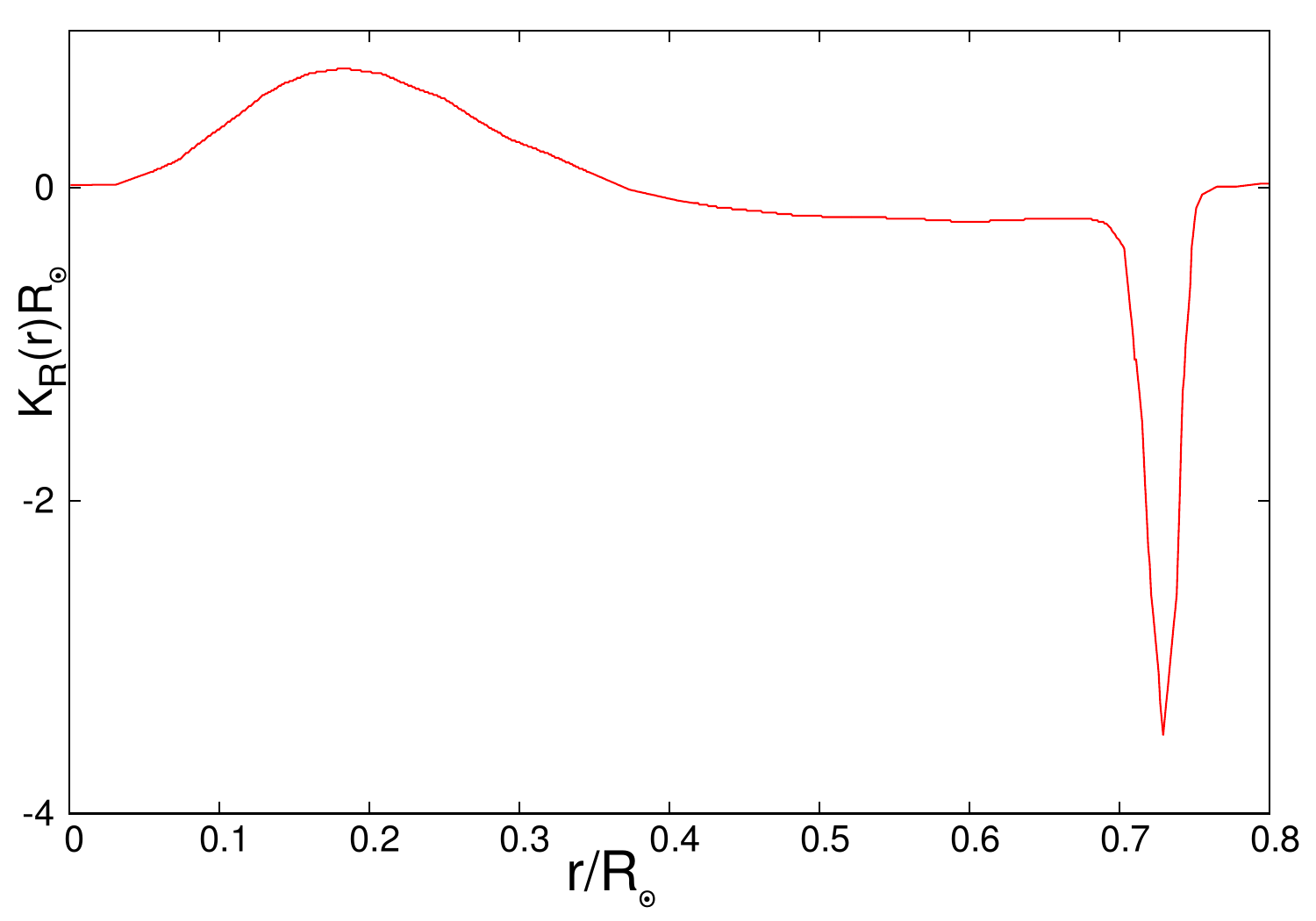}
\caption{Functional derivative $K_Y(r)$ of surface He abundance with respect to opacity.}
\vskip 1.0 cm
\label{fig4}
\end{figure}

The surface Helium abundance kernel, $K_Y(r)$, has been calculated in Villante 2010 and is plotted in Fig.~\ref{fig4}. It is important to notice that the kernel is positive-valued and, thus, the surface Helium abundance is highly sensitive to the overall scale of the opacity profile. Recall we discussed in Sec.~\ref{sec:opacity} how the scale of the vSZ16 $\delta \kappa (r)$ is higher than that of the ``ideal" variation with respect to the AGSS09 baseline model $\delta \kappa_{\text{id}}(r)$ (see Fig.~\ref{fig1}). Given the fact that the kernel $K_Y(r)$ is positive-valued, we expect that vSZ16 abundances will lead to a surface Helium abundance larger than that inferred by observations (recall instead that AGSS09 abundances predict a value for $Y_s$ which is too low by $\approx 7\%$).

We could have reached the above conclusion by a simpler heuristic argument. We already saw in Sec.~\ref{sec:opacity} that an increase in the abundance of refractories correlates with a hotter core. Increasing the temperature of the core would result in an increase in the nuclear reaction rates which in turn works to increase the luminosity of the Sun. However, the latter is very well measured and cannot be modified. In order to keep its luminosity fixed, the Sun responds by reducing its Hydrogen abundance $X$. However, given that $X + Y + Z = 1$, a decrease in $X$ has to correspond to an increase in the Helium abundance $Y$, and hence an increase in the surface Helium abundance $Y_s$ as well (see Vinyoles and Vogel 2016 for further discussions on the matter).

\subsubsection{Radius of convective zone boundary}\label{sec:czbobservables}

\begin{figure}[!h]
\includegraphics[width=.3\textwidth , angle=270]{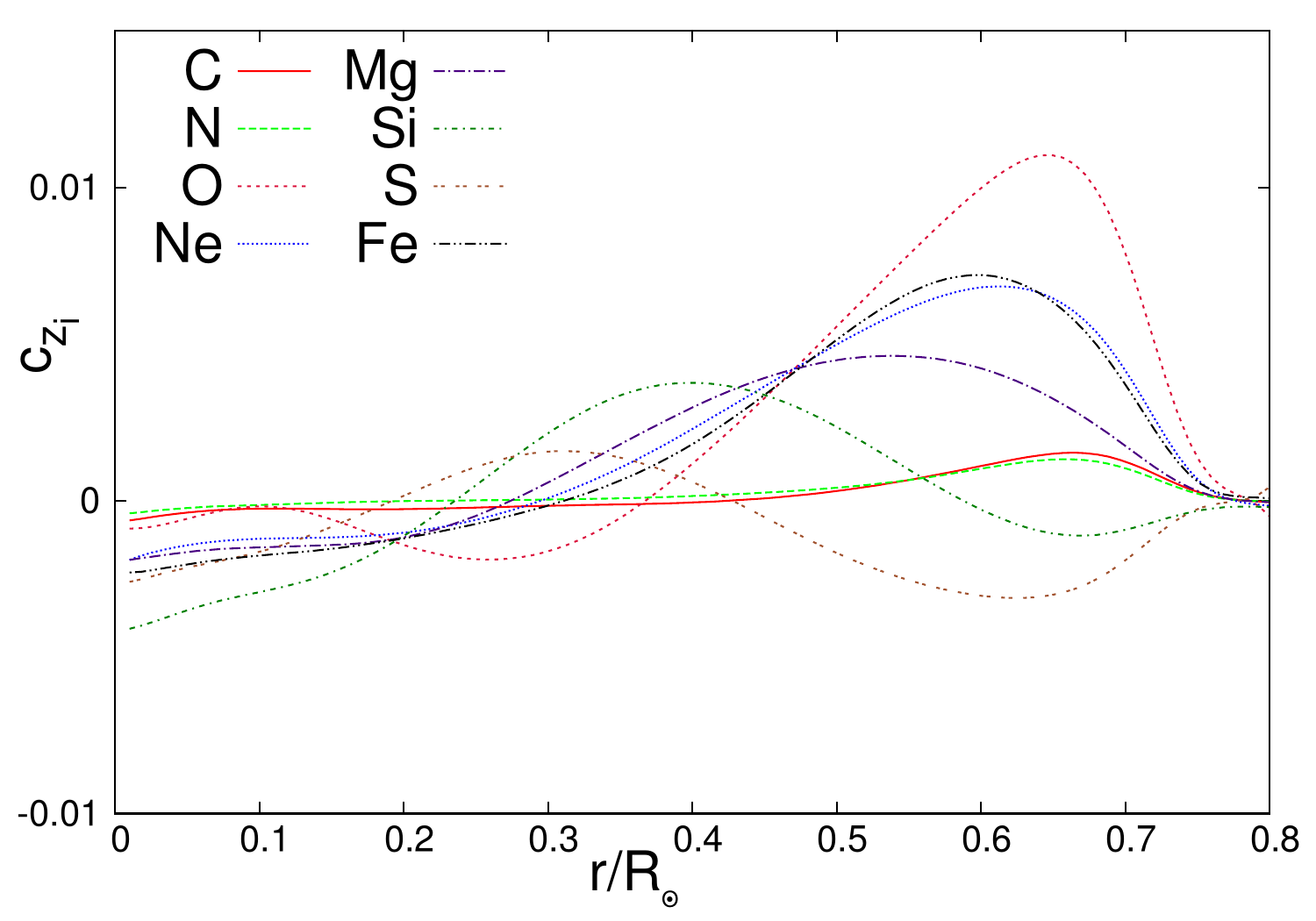}
\caption{Same as Fig. \ref{fig4} for convective radius, $K_R(r)$. We note that the convective radius is found at $r \sim 0.73 R_\odot$.}
\label{fig5}
\end{figure}

The radius of CZB kernel $K_R(r)$ has been worked out in Villante 2010, and is plotted in Fig.~\ref{fig5}. As with the sound speed, the radius of CZB too is very sensitive to the opacity profile at the base of the convective zone. This is the reason behind the sharp peak at $r \approx 0.73R_{\odot}$ in the radius of CZB kernel $K_R(r)$ (Fig.~\ref{fig5}).

Therefore, for the radius of CZB we can draw analogous conclusions as for the sound speed: because the radius of CZB is most sensitive to the abundance of volatiles (which in vSZ16 is closer to the previous concordance value of GS98 than those of AGSS09 are), we expect the location of the radius of CZB of vSZ16 to match observations better than that of AGSS09 (recall that the AGSS09 abundances predict a too shallow radius of CZB by $\approx 1.5\%$).

\subsubsection{Neutrino fluxes}\label{sec:neutrinosobservables}

Finally, we consider the following five neutrino fluxes: $\Phi_{\text{pp}}$, $\Phi_{\text{Be}}$, $\Phi_{\text{B}}$, $\Phi_{\text{N}}$, and $\Phi_{\text{O}}$. The neutrino kernels have been calculated in Villante 2010. Their main broad features are that they essentially drop off to zero for $r/R_{\odot} \gtrsim 0.45$, representing the well-known fact that neutrino fluxes are extremely sensitive to the conditions in the deep interior of the Sun. For the same reason, neutrino fluxes are extremely sensitive to the abundance of refractory elements, which play a major role in shaping the opacity profile near the core of our star. Of the five kernels, all but the one corresponding to the pp neutrino fluxes are positive-valued almost everywhere, reflecting the fact that an increase in opacity implies an increase in neutrino fluxes.

Instead of numerically integrating the neutrino kernels, we choose a more simple but equivalent method to estimate the variations in neutrino fluxes. Namely, the method of power-law exponents we already discussed previously. Given a certain neutrino flux $\Phi_i$ and power-law exponents for the given type of flux, $\varphi_{i,j}$ [notice the two different indices, $i$ running on the type of flux and $j$ running on the metals, i.e. $i=\text{pp},\text{Be},\text{B},\text{N},\text{O}$ and $j=\text{C},\text{N},\text{O},\text{Ne},\text{Mg},\text{Si},\text{S},\text{Fe}$], we can express the fractional variation in a given neutrino flux $\Phi_i$ as:
\begin{eqnarray}
\delta \Phi_i = \sum _j \varphi_{i,j}\delta Z_j
\label{dphi}
\end{eqnarray}
The values for $\varphi_{i,j}$ we adopt are taken from Villante et al. 2014, and are tabulated in Tab.~\ref{tab2}. Notice that, of course, both the fractional variations in fluxes $\delta \Phi_i$ and the power-law exponents $\varphi_{i,j}$ are dimensionless.

We see from Tab.~\ref{tab2} that neutrino fluxes are strongly sensitive to the abundance of refractories. The C and N neutrinos, for obvious reasons, are in addition strongly sensitive to the abundance of C and N (which are among the volatile elements instead). The discussion we held in Section \ref{sec:heliumobservables} for the surface Helium abundance will hold here as well. Namely, by virtue of the large abundance of refractory elements, we expect vSZ16 to lead to an overproduction of solar neutrinos. As we will see, the predicted fluxes will turn out to be well beyond the allowed limits of current measurements or upper limits.

\vskip 0.5 cm
\begin{table}[!h]
\caption{Power-law exponents relating variations in neutrino fluxes to variations in metal abundances. That is, the entry of the table in row $i$ and column $j$ correspond to $\varphi_{i,j}$, the logarithmic derivative of the $i$th neutrino flux with respect to the $j$th elemental abundance.}
\label{tab2}
\begin{center}{
\scriptsize
\begin{tabular}{|l|cccccccc|} 
\hline
\hline
$\downarrow i \ \ \ \rightarrow j$ & C & N & O & Ne & Mg & Si & S & Fe \\ \hline
$\text{pp}$ & -0.005 & -0.001 &  -0.004 & -0.004 & -0.004 &  -0.008  &  -0.006 &  -0.017 \\ 
$\text{Be}$ &  0.004 &  0.002  & 0.052 & 0.046  & 0.048 & 0.103 & 0.073 & 0.204 \\
$\text{B}$ & 0.026 & 0.007 & 0.112 & 0.088 & 0.089 & 0.191 & 0.134 &  0.501 \\
$\text{N}$ & 0.874 & 0.147 &  0.057 & 0.042 & 0.044 & 0.102  & 0.072 & 0.263 \\ 
$\text{O}$ &  0.827 & 0.206 & 0.084 & 0.062 & 0.065 & 0.145 & 0.102 & 0.382 \\ \hline
\end{tabular}
}\end{center}
\end{table}

\section{\textbf{Results}}\label{sec:results}

In this Section we present our results for the response of the helioseismology observables to the change in solar element abundances from the older results of AGSS09 to the new \textit{in situ} measurements of vSZ16.  We conclude with a discussion on caveats to the applicability of our methodology. 

\subsection{Sound speed}\label{sec:soundresults}

\begin{figure}[!htb]
\includegraphics[width=.35\textwidth,angle=270]{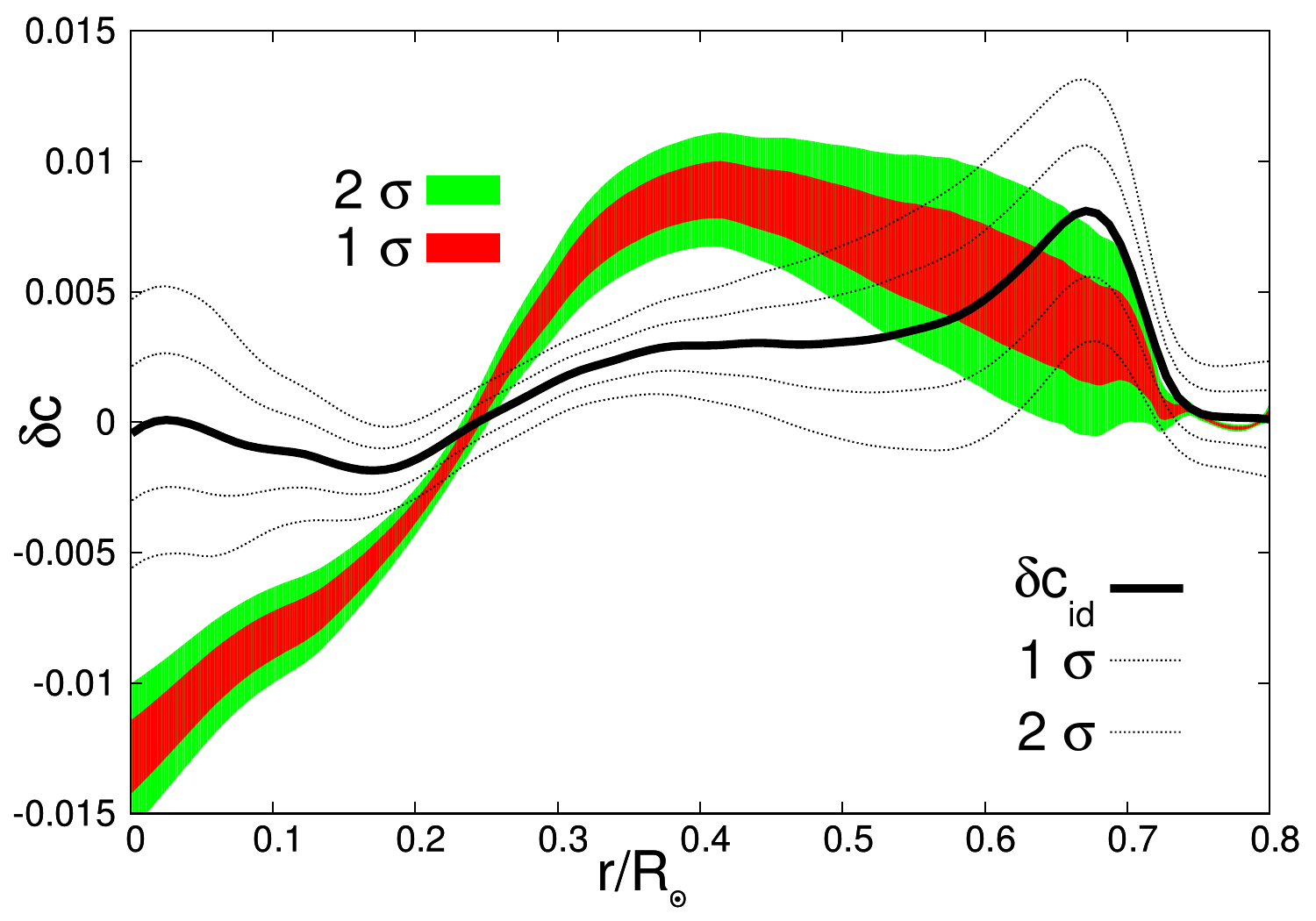}
\caption{Fractional variation in the sound speed of vSZ16 with respect to the baseline AGSS09 model, i.e. $\delta c(r)$ [for an operative definition see Eq.~(\ref{deltaur})], with 1$\sigma$ (red) and 2$\sigma$ (green) uncertainty bands propagated from the uncertainties on vSZ16 abundances [through Eq.~(\ref{deltaur})]. The thick solid line is $\delta c_{\text{id}}$ (variation which brings AGSS09 sound speed in agreement with helioseismology). The dotted lines represent 1$\sigma$ and 2$\sigma$ uncertainties on $\delta c_{\text{id}}$, obtained from the combination of solar model, statistical and systematic uncertainties in quadrature (see also Fig.~1 of Serenelli et al. 2016). The radius of CZB is located at $r/R_{\odot} \simeq 0.73$.}
\label{fig6}
\end{figure}

The results for the sound speed are presented in Fig.~\ref{fig6}. We also plot 1$\sigma$ and 2$\sigma$ (red and green respectively) error bands on $\delta c(r)$, obtained by propagating the uncertainties on $\delta Z_j$ through the logarithmic derivatives $c_{Z_i}$. The obtained response $\delta c(r)$ is to be compared with the thick solid line in the figure, which represents the fractional difference between the sound speed inferred from helioseismology and the sound speed in the baseline AGSS09 model. Therefore, the thick solid line corresponds to the fractional variation required to bring the AGSS09 sound speed in agreement with helioseismological inferences. We refer to this profile as $\delta c_{\text{id}}$. The uncertainty on $\delta c_{\text{id}}(r)$, denoted by the dotted lines, is the total uncertainty due to solar model, statistical uncertainty (coming from uncertainties in solar frequency measurements) and systematic uncertainties from the modelling procedure (this is the same error bar reported in Fig.~1 of Serenelli et al. 2016). For vSZ16 abundances to bring the sound speed in agreement with helioseismology, $\delta c(r) = \delta c_{\text{id}}(r)$ is required.

\begin{figure}[!h]
\includegraphics[width=.35\textwidth,angle=270]{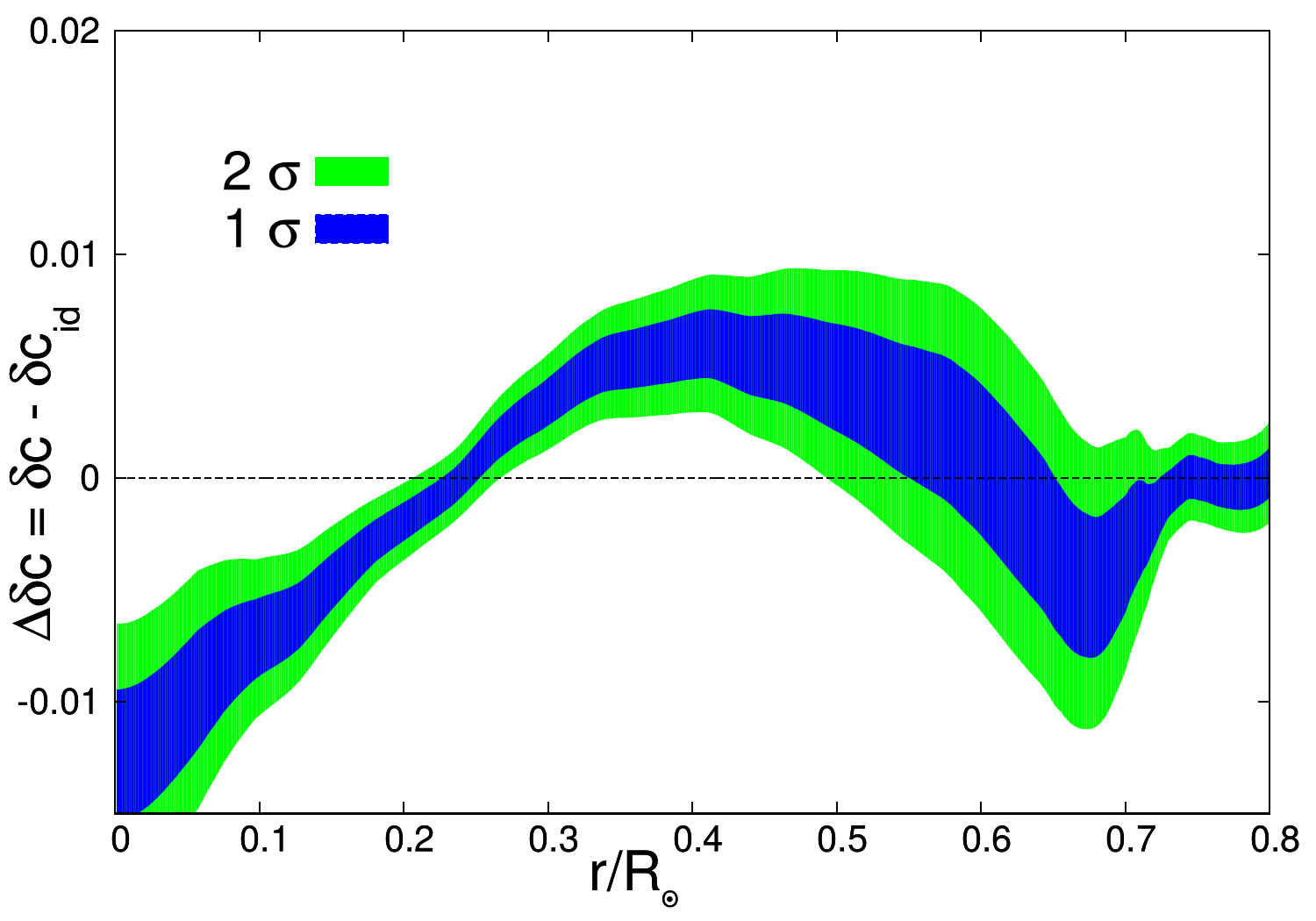}
\caption{Difference between $\delta c(r)$ and $\delta c_{\text{id}}(r)$, i.e. $\Xi(r)\equiv\delta c(r)-\delta c_{\text{id}}(r)$ [see Eq.~(\ref{deltadelta})], with 1$\sigma$ (blue) and 2$\sigma$ (green) uncertainty bands obtained through Eq.~(\ref{sigmaxi}). The quantity $\Xi(r)$ is to be compared with the dashed line at $\Xi = 0$ (that is, the x-axis), which would correspond to perfect agreement between the sound speed profile of the Sun and the sound speed obtained with the vSZ16 abundances. The radius of CZB is located at $r/R_{\odot} \simeq 0.73$.}
\vskip 0.5 cm
\label{fig7}
\end{figure}

A better visual comparison between $\delta c(r)$ and $\delta c_{\text{id}}(r)$ can be obtained instead if we plot the difference between the two, that is, the following quantity:
\begin{eqnarray}
\Xi(r) \equiv \delta c(r) - \delta c_{\text{id}}(r)
\label{deltadelta}
\end{eqnarray}
This is done in Fig.~\ref{fig7}. A perfect agreement between model and helioseismology then corresponds to $\Xi(r) = 0$ (the x-axis). The uncertainty on $\Xi(r)$ is obtained by combining the uncertainties on $\delta c(r)$ and $\delta c_{\text{id}}(r)$ in quadrature, given that the two are independent, that is:
\begin{eqnarray}
\sigma_{\Xi}(r) = \sqrt{\sigma_{\delta c}(r)^2+\sigma_{\delta c_{\text{id}}}(r)^2} \, .
\label{sigmaxi}
\end{eqnarray}
\newline

As we anticipated in Sec.~\ref{sec:soundobservables}, the vSZ16 sound speed profile represents an improvement over that of AGSS09 near the radius of CZB and at intermediate radii, where volatiles (and in particular C and O, whose values are quite close to those of GS98) play a major role in shaping the opacity profile. In particular, the discrepancy between vSZ16 and helioseismology at the radius of CZB is reduced to a mere $0.68\sigma$.\footnote{When assessing the degree of discrepancy between two values of the same observable ${\cal O}_1 \pm \sigma_{{\cal O}_1}$ and ${\cal O}_2 \pm \sigma_{{\cal O}_2}$, the number of $\sigma$s we quote is given by: $\vert {\cal O}_1-{\cal O}_2 \vert / \sqrt{\sigma_{{\cal O}_1}^2+\sigma_{{\cal O}_2}^2}$.} Above the radius of CZB, the disagreement between model and helioseismology essentially disappears because the temperature gradient becomes adiabatic (ensuing the breaking of hydrostatic equilibrium and causing convection to set in), and $c(r)$ depends no longer on the composition of the Sun. Our finding that vSZ16 represents an improvement over AGSS09 at intermediate and large radii agrees with the findings of Serenelli et al. 2016.

Closer to the center, vSZ16 instead fares considerably worse than AGSS09. This can be once more traced back to the huge increase in the abundance of refractory elements, which are mostly responsible for shaping the opacity profile near the core. In particular, near the core, the discrepancy between vSZ16 and helioseismology is at the level of $4.2\sigma$.

We can construct an ``effective" number of $\sigma$s representing the average deviation of the vSZ16 sound speed from helioseismology. To do so, we take $i=80$ equispaced couples of points $\{\delta c_i,\delta c_{\text{id},i}\}$ between $0 R_{\odot}$ and $0.8 R_{\odot}$ along $\delta c(r)$ and $\delta c_{\text{id}}(r)$ (or, equivalently, 80 equispaced points $\Xi_i$). Then, we compute the quantity (see footnote 3 for a mathematical justification behind this choice):
\begin{eqnarray}
\sigma_{\text{eff}} = \frac{1}{80}\sum_i \frac{\vert \delta c_i-\delta c_{\text{id},i} \vert}{\sqrt{\sigma_{\delta c_i}^2+\sigma_{\delta c_{\text{id},i}}^2}} = \frac{1}{80}\sum_i \frac{\vert \Xi_i \vert}{\sigma_{\Xi_i}} \, .
\label{discrete}
\end{eqnarray}
Using the definition in Eq.~(\ref{discrete}), we find $\sigma_{\text{eff}} \simeq 2.5$, confirming the fact that, despite the improvement over AGSS09 at intermediate and large radii, vSZ16 still disagrees by a large margin when compared to data from helioseismology. We can actually construct the continuous version of Eq.~(\ref{discrete}), namely:
\begin{eqnarray}
\sigma_{\text{eff}} &=& \frac{1}{r_1-r_2}\int_{r_1}^{r_2} dr \ \frac{\vert \delta c(r)-\delta c_{\text{id}}(r) \vert}{\sqrt{\sigma_{\delta c}(r)^2+\sigma_{\delta c_{\text{id}}}(r)^2}} \nonumber \\
&=& \frac{1}{r_1-r_2}\int_{r_1}^{r_2} dr \ \frac{\vert \Xi(r) \vert}{\sigma_{\Xi}(r)} \, ,
\label{continuous}
\end{eqnarray}
where $r_1=0R_{\odot}$ and $r_2\approx0.8R_{\odot}$. Using the continuous version given by Eq.(\ref{continuous}), we find once more $\sigma_{\text{eff}} \simeq 2.5$, confirming the disagreement between the vSZ16 sound speed and helioseismology data. We stress that the quantity $\sigma_{\text{eff}}$ just gives a broad quantification of the disagreement between vSZ16 and helioseismology, and is of limited statistical usefulness. It is in general more useful to refer to the disagreement between the two at a given radius $r$, rather than consider the average of the latter figure.

\subsection{Surface Helium abundance}\label{sec:heliumresults}

We compute the variation in the surface Helium abundance using the methodology described in Sec.~\ref{sec:heliumobservables}, and find an absolute variation of $\Delta Y_s=0.052 \pm 0.025$, where the uncertainty has been obtained propagating the uncertainties on the vSZ16 abundances through the relevant power-law exponents.

The Standard Solar Model implemented with AGSS09 abundances predicts a value $Y_{s,\text{AGSS09}}=0.232 \pm 0.003$, whereas the value inferred from helioseismology is $Y_{s,h}=0.2485 \pm 0.0034$. Using the value of $Y_{s,\text{AGSS09}}$ and the value of $\Delta Y_s$ we found in our analysis, we infer the value of the vSZ16 surface Helium abundance through $Y_{s,\text{vSZ16}}=Y_{s,\text{AGSS09}}+\Delta Y_s$, obtaining $Y_{s,\text{vSZ16}}=0.284 \pm 0.025$. As we see, the central value of the surface Helium abundance predicted by vSZ16 is significantly larger than that inferred from helioseismology, just as we had anticipated in Sec.~\ref{sec:heliumobservables} on the basis of the observation that the surface Helium abundance is very sensitive to the abundance of refractory elements.

Because of the large uncertainty on $Y_{s,\text{vSZ16}}$ (an order of magnitude larger than that on $Y_{s,\text{AGSS09}}$), a quantification of the disagreement between vSZ16 and helioseismology and a comparison with the disagremeent between AGSS09 and the latter is not appropriate.\footnote{If we were to go ahead and compute the number of $\sigma$s of discrepancy between $Y_{s,\text{vSZ16}}$ and helioseismology as $\vert Y_{s,\text{vSZ16}}-Y_{s,h} \vert / \sqrt{\sigma_{Y_{s,\text{vSZ16}}}^2+\sigma_{Y_{s,h}}^2}$ (see footnote 3), we would obtain $1.4$, which of course is a poor representation of the true situation, given the large uncertainty on $Y_{s,\text{vSZ16}}$ (an order of magnitude larger than that on $Y_{s,\text{AGSS09}}$).} Instead, we conclude that vSZ16 abundances do not solve the surface Helium abundance problem, and actually aggravate the issue. Our results on the surface Helium abundance agree with those of Serenelli et al. 2016.

\subsection{Convective radius}\label{sec:czbresults}

We compute the variation in the radius of CZB using the method described in Sec.~\ref{sec:czbobservables}, and obtain $\delta R_b=-0.011 \pm 0.004$, where once more the uncertainty has been obtained propagating the uncertainties on the vSZ16 abundances through the relevant power-law exponents.

The Standard Solar Model implemented with AGSS09 abundances predicts a value $R_{b,\text{AGSS09}}=0.723 \pm 0.002$, whereas the value inferred from helioseismology is $R_{b,h}=0.713 \pm 0.001$. Using the value of $R_{b,\text{AGSS09}}$ and the value of $\delta R_b$ we found in our analysis, we infer the value of the vSZ16 radius of CZB through $R_{b,\text{vSZ16}}=R_{b,\text{AGSS09}}(1+\delta R_b)$, obtaining $R_{b,\text{vSZ16}}=0.715 \pm 0.002$. The discrepancy between vSZ16 and helioseismology is reduced to the level of $0.88\sigma$ (see footnote 3).

We conclude that the VSZ16 abundances greatly alleviate the radius of CZB problem. Our conclusion, which agrees with that of Serenelli et al. 2016, had already been reached in Sec.~\ref{sec:czbobservables} on the basis that the radius of CZB is mostly sensitive to the abundance of volatiles rather than refractories.

\subsection{Neutrino fluxes}\label{sec:neutrinosresults}

We use the method described in Sec.~\ref{sec:neutrinosobservables} to compute the response of Solar neutrinos to the vSZ16 abundances. We find fractional variations given by: $\delta \Phi_{\text{pp}} = -0.038 \pm 0.004$, $\delta \Phi_{\text{Be}} = 0.42 \pm 0.05$, $\delta \Phi_{\text{B}} = 0.88 \pm 0.10$, $\delta \Phi_{\text{N}} = 1.09 \pm 0.15$, $\delta \Phi_{\text{O}} = 1.27 \pm 0.15$. We stress that these values are only determined at linear order and, particularly for the N and O neutrinos, second-order effects are likely to be playing a role (we discuss further in Sec.~\ref{sec:applicabilityoflsm}). Nonetheless, our linear analysis brings us to conclude that neutrino fluxes with vSZ16 abundances are in severe disagreement with observations.

The above variations imply that the pp neutrinos would be slightly overproduced compared to current bounds (see e.g. Table 2 in Serenelli et al. 2016), whereas Be and B neutrinos are severely overproduced (by up to a factor of 2). We are still lacking a detection of CNO neutrinos, although the SNO$^+$ collaboration (Andringa et al. 2016) could possibly do it within the next years. Currently we only have very rough upper limits on N and O neutrinos from Borexino (Bellini et al. 2010; Bellini et al. 2011; Bellini et al. 2014). The increase in N neutrinos predicted by vSZ16 is still marginally allowed within these upper limits, whereas the increase in O neutrinos is excluded.

We conclude that vSZ16 abundances are in extremely strong tension with the very accurate Be and B neutrino flux measurements, and in less severe tension with the upper limits on N and O neutrinos. The tension is once more due to the large variations in the abundances of refractory elements, which entail a hotter core and therefore an overproduction of neutrinos. Our findings agree with those of Serenelli et al. 2016.

\subsection{On the applicability of the LSM}\label{sec:applicabilityoflsm}

We end the discussion with a caveat on the applicability of our methodology. We have performed a first-order analysis based on the Linear Solar Model. Strictly speaking, this method is only valid for opacity variations sufficiently smaller than unity, that is, $\delta \kappa < {\cal O}(1)$. Yet, at intermediate radii, $\delta \kappa \approx 0.5$, so that we might expect the above approximation to break down and second-order effects might alter some of our results. Our formalism does not capture higher-order effects, which can only be treated by doing a full non-linear study using e.g. solar codes. This, in fact, is the more complete approach taken in Serenelli et al. 2016. From a practical point of view, however, we notice a very good agreement between our results and those of Serenelli et al. 2016. This lends us confidence \textit{a posteriori} with regards to the goodness of our analysis and the applicability of the LSM to our problem (despite the above concern).

We can, however, make a more compelling case for the validity of the LSM to the problem we are considering. As we have discussed at length in this work, the observables which are most sensitive to the abundance of volatiles (that is, sound speed and radius of CZB) are also most sensitive to the opacity at the bottom of the radius of CZB. Conversely, the observables which are most sensitive to the abundance of refractories (that is, surface Helium abundance and neutrino fluxes) are also most sensitive to the opacity in the core of the Sun. We notice that, both in the core of the Sun and at the radius of CZB, the central value of the variation in opacity is of order $\delta \kappa \approx 0.25-0.35$ (see Fig.~\ref{fig1}), which is sufficiently small that the linear approximation might still be valid. We conclude that the good agreement between the results of the LSM and those obtained from the non-linear analysis of Serenelli et al. 2016 can be traced to the fact that the observables we considered are mostly sensitive to regions of the Sun where the opacity variation $\delta \kappa$ is sufficiently small. One should, however, always keep these caveats in mind when comparing our results with those of Serenelli et al. 2016.

In addition, it is known that the the Sun responds linearly to relatively large opacity variations, a fact which had first been noticed in Tripathy \& Christensen-Dalsgaard 1998. The reason for this unexpected behaviour is not completely understood. However, it is likely to be connected to the fact that, when large opacity variations are considered, some of the initial parameters of solar codes (e.g. the initial Helium fraction) need to be adjusted in order to satisfy the observational constraints on the Sun's radius and luminosity. While these adjustments are automatically taken care of in the LSM, their physical effects is to partially reduce the initially large opacity variations to which the Sun is subject (Francesco Villante, personal communication). In any case, the net result is that, despite the relatively large variations in opacity, the LSM is applicable to our study.

Our results for neutrino fluxes deserve an aside. Since their variations in response to the change in abundance from AGSS09 to vSZ16 has been estimated to be of ${\cal O}(1)$ at linear order, we expect second-order effects to impact them the most. The LSM alone is not able to provide us the direction in which these effects go (that is, they might in principle go in the opposite direction with respect to the first-order ones and hence possibly ameliorate the tension). However, we can cross-check our results with the full non-linear study of Serenelli et al. 2016. The conclusion is that second-order effects go in the same direction as the first-order ones, namely, they act to further increase the neutrino fluxes and hence worsen the disagreement with observations.

It is worth remarking once more that our linear analysis reaches the same conclusion as the full non-linear study of Serenelli et al. 2016: namely, that vSZ16 abundances do not solve the ``solar modelling problem", thus reaching our goal of determining whether solar wind measurements are able to solve this long-standing issue. Moreover, our work also serves to highlight the goodness of the LSM as a tool to analyze the solar interior in a simple and transparent way.

\section{\textbf{Discussion and conclusion}}\label{sec:conclusion}

This work has a simple and straightforward purpose: it is to assess the implications for solar models of the abundances provided by von Steiger \& Zurbuchen 2016 through solar wind analysis. For that purpose, we conducted a first-order analysis of the response of helioseismological observables to the change in abundances with respect to the previous widely used set by Asplund et al. 2009. Our results indicate that, whereas for the convective zone boundary the overall agreement between solar data and the predicted behaviours is increased, the disagreement with the surface Helium abundance is instead considerably worsened. The sound speed predicted by vSZ16 is considerably improved over that of AGSS09 at intermediate and large radii, but the discrepancy is severely worsened at small radii. The predictions for neutrino fluxes are strongly discrepant with current measurements:  the Be and B neutrino fluxes predicted by vSZ16 are too high by up to a factor of 2. Our overall conclusion is that vSZ16 abundances studied in the Linear Solar Model do not solve the ``solar modelling problem".

We have identified the physical reason underlying both the improved agreements and worsened disagreements. On the one hand, the increase in volatile abundances (especially C and O) has brought their values closer to the previous concordance values of Grevesse \& Sauval 1998. Volatile elements play a dominant role around the convective zone boundary, and thus their increase in abundance improves the agreement of observables which are most sensitive to the opacity profile in that region: the sound speed profile and the convective zone boundary. On the other hand, the very large increase in the abundance of refractories, in particular Si and S, correlates with an increase in core temperature. Thus, the excessive increase in abundance of refractories worsens the disagreement of observables which are very sensitive to the conditions of the core, that is, surface Helium abundance and neutrino fluxes.

Obviously, the vSZ16 data themselves do not appear to address the ``solar modelling problem". That might be due to residual fractionation in the solar atmosphere, which is not excluded by vSZ16 and also suggested by Serenelli et al. 2016, and which would make solar wind abundances an unreliable estimate of the bulk solar chemical composition, unless the associated systematics are taken into account. In fact, the ratio between abundances in the vSZ16 and AGSS09 catalogues shows a remarkable correlation with the first ionization potential (FIP) of the elements in question. This suggests that FIP fractionation is playing an important role, feeding additional systematics into solar wind measurements. It is also worth noticing that FIP fractionation appears to increase or decrease the measured abundance of elements depending on whether their FIP potential is greater or smaller than that of Hydrogen, which could explain the large increase in the abundance of refractory elements. On this note, it may be worth going back and taking a close look at both remote and \textit{in situ} measurements of refractory elements (see e.g. Landi et al. 2012). If FIP fractionation is indeed playing an important role, this would also invalidate the argument for which the \textit{in situ} measured metallicity represents a lower limit to the true metallicity of the Sun (which relied on presence of residual fractionation processes which only decreased but not increased the inferred metallicity).

Finally, there might indeed be important physical mechanisms at play that remain to be discovered. It would not be the first time the Sun and its composition tell us something fundamental about physics!

\acknowledgments

K.F. acknowledges  support  from  DoE  grant  de-sc0007859 at  the  University  of  Michigan  as  well  as  support  from
the Michigan Center for Theoretical Physics. KF and SV are supported by the Vetenskapsr\r{a}det (Swedish Research Council) through contract No.  638-2013-8993 and the Oskar Klein Centre for Cosmoparticle Physics. THZ is partially supported by NASA NNX13AH66G. SV thanks the Michigan Center for Theoretical Physics for hospitality while this work was conducted. THZ acknowledges the International Space Science Institute in Bern, where much of the analysis relevant to this work has been performed. SV thanks Mads Frandsen, Subir Sarkar, Pat Scott, Ian Shoemaker, and Francesco Villante for useful discussions and correspondence. We also thank Enrico Landi for constructive feedback on a previous version of the paper. We thank the anonymous referee for useful remarks which helped improve our paper. \newline

\textit{Note added}: In response to the first draft of this paper (Vagnozzi et al. 2016), Serenelli et al. 2016 wrote a paper which criticized our main findings and prompted us to repeat our analysis. We are extremely grateful to them for prompting us to a more careful analysis, which has led us to conclude that vSZ16 does not solve the ``solar modelling problem".

\end{document}